# USING DATA WAREHOUSE TO SUPPORT BUILDING STRATEGY OR FORECAST BUSINESS TEND

Phuc V. Nguyen, M.Sc., Ph.D. (candidate)
Faculty of Information Technology, Hung Vuong Univesity, Ho Chi Minh City, Vietnam
Phone: (+84) 938.20.11.02 - Email: x201102x@gmail.com

**ABSTRACT**

The data warehousing is becoming increasingly important in terms of strategic decision making through their capacity to integrate heterogeneous data from multiple information sources in a common storage space, for querying and analysis. So it can evolve into a multi-tier structure where parts of the organization take information from the main data warehouse into their own systems. These may include analysis databases or dependent data marts. As the data warehouse evolves and the organization gets better at capturing information on all interactions with the customer. Data warehouse can track customer interactions over the whole of the customer's lifetime.

**Keywords**

Data, warehouse, lifecycle, CRM, Decision-makers, Data marts, business, intelligence, olap, etl,

## 1. INTRODUCTION

Data warehouse is a heart of Business Intelligence which is essential for any effective application. In other words, data warehouse is a consolidated view of your enterprise data, optimized for reporting and analysis. Hence, data warehouse can greatly enhance abilities of decision making. In terms of CRM, data warehouse and data mining are utilized to provide valuable reports about actual business as well as to forecast about needs of customer. We will discuss more about characteristics of data warehouse and how it is reflected in the project.

Actually, the company does not have anything using data warehouse to support building strategy or forecast business tend. All the jobs of data collection and consolidation have been done manually. To improve the performance of the tasks, the company should own a methodology and data warehouse infrastructure:

### 1.1 Data warehouse lifecycle

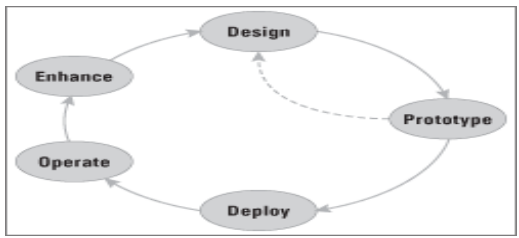

Figure 1: Data warehouse lifecycle

The DW lifecycle shows that we must pay more attention at Design and Prototype steps before implementing (Deploy step) to ensure that result will match with user desire. This is very important to lead to success of a data warehouse project. After go-live (Operate step), we must continue to enhance it to gain more effective. This is a careful way of business development.

### 1.2 Data warehouse environment

With changes of IS in S&M department, the number of information sources has increased one more. Therefore, we must need to unify the source by copying data to from Operational applications to one place, called Staging area. The technique is ETL, Extract – Transform - Load, an important technique in data warehouse. From the Staging DB, data will be cleaned up and copied to Data Warehouse DB by using ETL technique as well and finally from Data Warehouse to a set of conformed Data Marts which are accessible by decision makers.

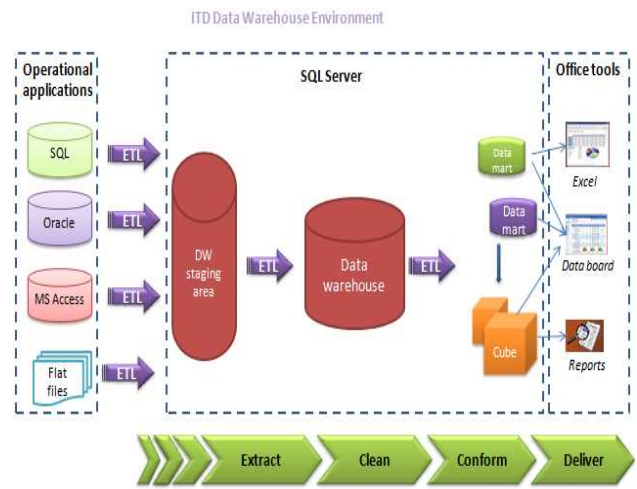

Figure 2: Data warehouse environment

### 1.3 Data warehouse architecture

Corresponding to the above environment, a corresponding architecture is below:

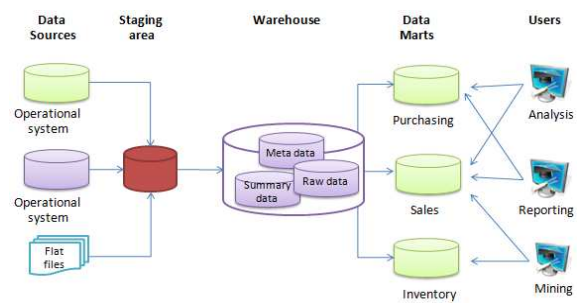

Figure 3: Data warehouse architecture

In the architecture, the data warehouse includes types of data like Meta data, raw data and Summary data. From the DW, meaningful data with new view will be generated to Data Marts for Purchasing, Sales & Marketing and Inventory. Decision-makers will access the data marts to forecast and to give right decision. It also allows generating meaningful reports instead of doing manually as before. This saves time a lot. "*Save time is save money*"

## 1.4 Data warehouse security

DW has become the most significant repository for business intelligence and decision-making support. Thus, the system security is set at a crucial priority. The company implements highest level of security by establishing user profiles (user groups) with different of authority concerning to the accessible information, performed operation, etc… and requiring regularly updated password. This will be defined more detail in section **Operation and Policies.**

The below image shows overview of data accessibilities:

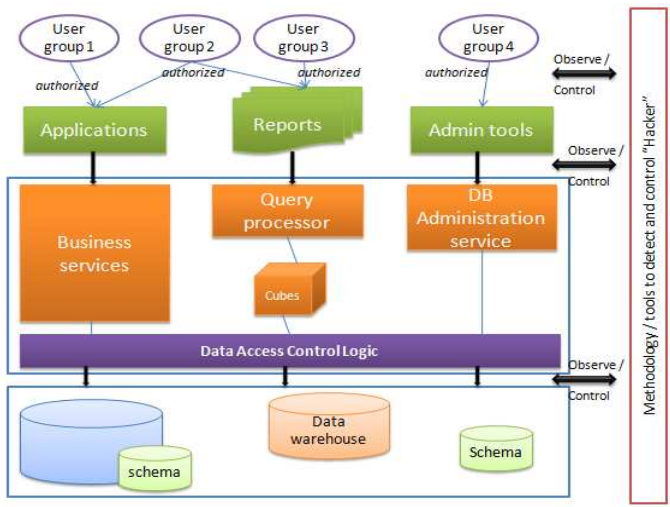

Figure 4: DB security model

## 1.5 Data warehouse framework and DBMS

In terms of DBMS, we choose MS SQL Server to deploy the project because it provides benefits like:

- Well-supported if bugs are reported
- Provide ability of data encryption on DB server itself
- It is also Microsoft technology so we can apply security policies across systems in the domain
- Provide a strong framework with powerful components for data warehouse. The below image illustrates a workflow from Design to movement of row data at Operational Sources to Data Transform/Data cleansing to Data Marts and to end-user tools.

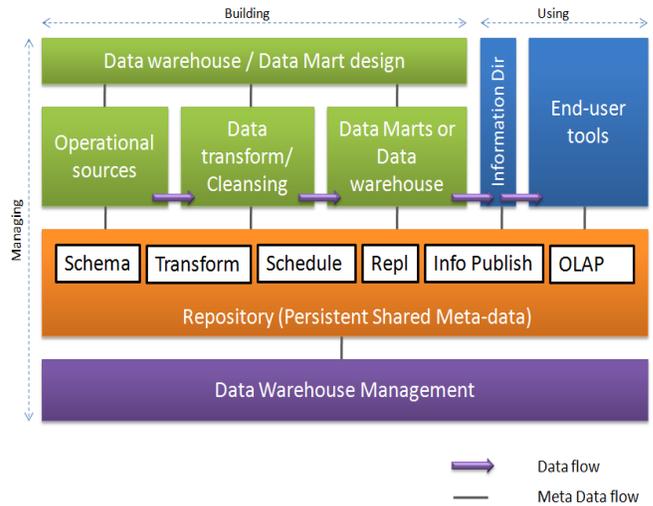

Figure 5: Microsoft Data warehouse framework

## 1.6 Conclusion

Data warehouse is designed to centralize different data sources. It creates a transparent data environment for user, helping user do data mining, report and analyze data efficiently. The DW would be applied in the future, at the higher level which already planned as mentioned.

## 2. ENTERPRISE DATA WAREHOUSE

This solution is common to develop enterprise data warehouse. Company is required more data and greater integration of data. Such as call center, customer communication, general ledger, project, human resources and operations. The volume of data expands and the complexity increases, this may result in many databases and data marts. Therefore, it is much more logical and beneficial to have one repository for data.

The data warehouse can evolve into a multi-tier structure where parts of the organization take information from the main data warehouse into their own systems. These may include analysis databases or dependent data marts. As the data warehouse evolves and the organization gets better at capturing information on all interactions with the customer. Data warehouse can track customer interactions over the whole of the customer's lifetime.

|  | Advantage | Disadvantage |
|---|---|---|
| Business | - Stops complex data analysis from interfering with normal business activity<br>- The company can refer to one 'single version of the truth' which can then feed numerous data marts with consistent data.<br>- Business can utilize existing customer information that systematized for their decision<br>- Customer information is managed effectively and fully by many data marts | - Business may not be able wait for the data warehouse to be implemented, it needs to make decisions today and a cheaper and less appropriate solution may be adopted and accepted.<br>- Costly |
| Technology | - Information in the data warehouse changes only periodically allowing meaningful comparisons to be made on stable sets of data which exist in between updates of the data warehouse.<br>- Databases were used for analysis, analyses made at different times would produce different results<br>- Company can utilize existing applications and IS infrastructure | - Take time for deployment<br>- Enterprise data warehouses are large and complex IT systems that require significant investment.<br>- Company needs to have professional resources to maintain and manage system<br>- One data marts outage may affect whole system |

Figure 6: Advantage and Disadvantage of Business and Technology

At this level, we can manage customer information perfectly. EDW integrates Customer data warehouse, Exchange server, Call accounting and EFFECT system, beside that, the company needs deploy more other application or enhance existing application. This is required a complex IS infrastructure and human resource for deployment. But, company can earn form them many benefit. Customer information is gathered multiplex allows the company analyze and come out decision exactly. Building an enterprise data warehouse can be a complex. Some organizations prefer to build smaller data marts first and let their data warehouse evolve over time. However, an enterprise data warehouse can prove essential to the success

## 2.1 Building a Better Data Warehouse and BI Environment

Information is a company's best asset and today's leading businesses are harnessing it to make real-time critical business decisions that will impact the company's bottom line, meet regulatory deadlines and address customer needs. But, how can companies be assured that the information is always accurate and up to date with data being integrated from disparate sources and with different standards?

IT organizations are required to tackle these data management issues head on and solve the complex issues around designing the infrastructure, ensuring data quality, understanding the source of data and business rules, managing the metadata and extracting and transforming it to standardized formats for a data warehouse. Yet, solving these issues can take many weeks and months while your business users wait impatiently for this critical information.

## 2.2 The Challenges of Using Data Warehouses

As today's decisions in the business world become more real-time, the systems that support those decisions need to keep up. It is only natural that Data Warehouse, Business Intelligence, Decision Support, and OLAP systems quickly begin to incorporate real-time data. Data warehouses and business intelligence applications are designed to answer exactly the types of questions that users would like to pose against real-time data. They are able to analyze vast quantities of data over time, to determine what is the best offer to make to a customer, or to identify potentially fraudulent, illegal, or suspicious activity. Ad-hoc reporting is made easy using today's advanced OLAP tools. All that needs to be done is to make these existing systems and applications work off real-time data. The following are a few of the challenges of adding real-time data to these systems:

## 2.3 Enabling Real-time ETL

One of the most difficult parts of building any data warehouse is the process of extracting, transforming, cleansing, and loading the data from the source system. Performing ETL of data in real-time introduces additional challenges. Almost all ETL tools and systems, whether based on off-the-shelf products or custom-coded, operate in a batch mode. They assume that the data becomes available as some sort of extract file on a certain schedule, usually nightly, weekly, or monthly. Then the system transforms and cleanses the data and loads it into the data warehouse.

This process typically involves downtime of the data warehouse, so no users are able to access it while the load takes place. Since these loads are usually performed late at night, this scheduled downtime typically does not inconvenience many users. When loading data continuously in real-time, there can't be any system downtime. The heaviest periods in terms of data warehouse usage may very well coincide with the peak periods of incoming data. The requirements for continuous updates with no warehouse downtime are generally inconsistent with traditional ETL tools and systems. Fortunately, there are new tools on the market that specialize in real-time ETL and data loading. There are also ways of modifying existing ETL systems to perform real-time or near real-time warehouse loading.

## 2.4 Modeling Real-time Fact Tables

The introduction of real-time data into an existing data warehouse, or the modeling of real-time data for a new data warehouse brings up some interesting data modeling issues. For instance, a warehouse that has all of its data aggregated at various levels based on a time dimension needs to consider the possibility that the aggregated information may be out of synch with the real-time data. Also some metrics such as month-to-date and week-to-date may behave strangely with a partial day of data that continuously changes. The main issue regarding modeling however revolves around where the real-time data is stored, and how best to link it into the rest of the data model.

## 2.5 OLAP Queries vs. Changing Data

OLAP and Query tools were designed to operate on top of unchanging, static historical data. Since they assume that the underlying data is not changing, they don't take any precautions to ensure that the results they produce are not negatively influenced by data changes concurrent to query execution. In some cases, this can lead to inconsistent and confusing query results.

Relational OLAP tools are particularly sensitive to this problem because they perform all but the simplest data analysis operations by issuing multi-pass SQL. A multi-pass SQL statement is made up of many smaller SQL statements that sequentially operate on a set of temporary tables.

This presents two problems. The first problem is that the results of a query that takes even one minute are arguably not exactly real-time anymore. While this data latency may be acceptable to a retail division manager, it might not be ok for an application that is looking for atmospheric trends that indicate the presence of a tornado, or for an application detecting real-time credit card or telecommunications fraud.

The second problem is that given the multiple passes of SQL required to perform almost any relational OLAP reporting or analytical operation, any real-time warehouse is likely to suffer from the result set internal inconsistency issue discussed above. There's nothing like the numbers not adding up properly to make a user skeptical of a report. For more complex product affinity or trend detection analytics, the results may be so confusing as to be completely useless.

## 2.6 Scalability & Query Contention

The issue of query contention and scalability is the most difficult issue facing organizations deploying real-time data warehouse solutions. Data warehouses were separated from transactional systems in the first place because the type of complex analytical queries run against warehouses don't "play well" with lots of simultaneous inserts, updates, or deletes.

Usually the scalability of data warehouse and OLAP solutions is a direct function of the amount of data being queried and the number of users simultaneously running queries. Given a fixed amount of data, the number of users on the system is proportional to query response time. Lots of concurrent usages cause reports to take longer to execute.

While this is still true in a real-time system, the additional burden of continuously loading and updating data further strains system resources. Unfortunately the additional burden of a continuous data load is not just equivalent to one or two additional simultaneously querying users due to the contention between data inserts and typical OLAP select statements. While it depends on the database, the contention between complex selects and continuous inserts tends to severely limit scalability. Surprisingly quickly the continuous data loading process may become blocked, or what used to be fast queries may begin to take intolerably long to return.

## 2.7    Real-time Alerting

Most alerting applications associated with data warehouses to date have been mainly used to distribute email versions of reports after the nightly data warehouse load. The availability of real-time data in a data warehouse makes alerting applications much more appealing, as users can be alerted to real-time conditions as they occur in the warehouse, not just on a nightly basis.

The availability of real-time data makes products such as Micro Strategy's NarrowCaster and similar products from Cognos and Business Objects very valuable. But real-time alerting using these products brings its own set of challenges, as surprisingly these products, like many query tools from the same vendors, were not designed to operate on or tested against real-time data feeds.

These products operate on a schedule or event basis, so they can either trigger an alert every few minutes or hours, or need to be triggered by an external system. Solutions to address these challenges are available in the market. There is also the issue of threshold management. When alerts are triggered frequently, there needs to be a mechanism in place to make sure that once an alert is sent due to a condition in the warehouse that the alert is not continuously sent over and over again during each alerting cycle.

As we have seen in this article, real-time data warehousing and OLAP are possible using today's technology, but challenges lurk seemingly every step of the way. For the determined team armed with the right knowledge and experience, it is possible to make real-time reporting, analysis, and alerting systems work. The challenge is to make the right tradeoffs along the way, to make sure the systems meet the needs of the user base while ensuring that they don't collapse under their own weight, or cause existing production warehouses to malfunction

It is likely that a lot of the challenges discussed above will become less challenging over time, as database, ETL, OLAP, reporting, and alerting tool vendors begin to add features to their systems to make them work better with real-time data streams. In the meantime, it is important to make sure real-time warehousing systems are well planned and designed, and thoroughly tested under realistic data and user load conditions before they are deployed.

The benefits of data warehousing in real-time are becoming clearer every day. With the right tools, designs, advice, approaches, and in some cases tricks, real-time data warehousing is possible using today's technologies, and will only become easier in the future.

## 2.8    A combination between Data warehouse and ERP

At the same time, many companies have been instituting enterprise resource planning (ERP) software to coordinate the common functions of an enterprise. ERP software usually has a central database as its hub, allowing applications to share and reuse data more efficiently than previously permitted by separate applications. The use of ERP has led to an explosion in source data capture, and the existence of a central ERP database has created the opportunity to develop enterprise data warehouses for manipulating that data for analysis. This paper will provide an overview of the issues and challenges that the intersection of these two IS concepts are creating.

Data warehouses are one of the foundations of the decision support systems of many IS operations. They serve as the storage facility of millions of transactions, formatted to allow analysis and comparison. As defined by the "father of data warehouse", William H. Inmon, a data warehouse is "a collection of integrated, subject-oriented databases where each unit of data is specific to some period of time. Data Warehouses can contain detailed data, lightly summarized data and highly summarized data, all formatted for analysis and decision support" ("Building a Data Warehouse", Inmon, W. H.; Wiley, 1996). In the "Data Warehouse Toolkit", Ralph Kimball gives a more succinct definition: "a copy of transaction data specifically structured for query and analysis" ("The Data Warehouse Toolkit", Kimball, R.; Wiley, 2000). Both definitions stress the data warehouse's analysis focus, and highlight the historical nature of the data found in a data warehouse.

Enterprise Resource Planning software is a recent addition to the manufacturing and information systems that have been designed to organize the flow of data from process start to finish. This flow of information has existed since the first manufacturers traded with the first merchants, but until the advent of ERP software and the processes that accompany it; this information was largely ignored and not captured. ERP software attempts to link all internal company processes into a common set of applications that share a common database. It is the common database that allows an ERP system to serve as a source for a robust data warehouse that can support sophisticated decision support and analysis.

ERP software is divided into functional areas of operation; each functional area consists of a variety of business processes. The main, common functional areas of operation in most companies would include: Marketing and Sales; Production and Operations (Materials Management, Inventory, etc.); Accounting and Finance; Human Resources. Historically, businesses have had clear divisions among each of these areas, and IS development was also clearly delineated so that systems did not share data or processes and cross-functional analysis of information was not possible. Since all functional areas ARE interdependent, this separation was not a valid representation of a business' activities and the divisions among the many information systems created artificial barriers that needed to be overcome.

ERP software was designed to eliminate the barriers to sharing data and processes that occur when companies design and implement information systems for a single function or activity. ERP software coordinates the entire business process, and stores all the captured data in a common database, accessible to all the integrated applications of the ERP suite. As explained in "Concepts in Enterprise Resource Planning" (Brady, Monk, Wagner; Course Technology, 2001) companies can achieve many cost savings and related benefits from the use of ERP for transaction processing and management reporting through the use of the ERP's common database and integrated management reporting tools.

However, much of the work performed by managers and knowledge workers in the 21st century is not transaction or management reporting-based. The main activity of knowledge and management staff is analysis, and this analysis is supported by the development and use of decision support systems. The most common application of DSS in companies today is the data warehouse. With the use of the ERP's common database and the implementation of DSS/DW user support products companies can design a decision support/data warehouse database that allows cross-functional area analysis and comparisons for better decision-making.

Since companies usually implement an ERP in addition to their current applications, the problem of data integration from the various sources of data to the data warehouse becomes an issue. Actually, the existence of multiple data sources is a problem with ERP implementation regardless of whether a company plans to develop a data warehouse; this issue must be addressed and resolved at the ERP project initiation phase to avoid serious complications from multiple sources of data for analysis and transaction activity. In data warehouse development, data is usually targeted from the most reliable or stable source system and moved into the data warehouse database as needed. Identification of the correct source system is essential for any data warehouse development, and is even more critical in a data warehouse that includes an ERP along with more traditional transaction systems. Integration of data from multiple sources (ERP database and others) requires rigorous attention to the metadata and business logic that populated the source data elements, so that the "right" source is chosen.

Another troubling issue with ERP data is the need for historical data within the enterprise's data warehouse. Traditionally, the enterprise data warehouse needs historical data (see Inmon's definition). And traditionally ERP technology does not store historical data, at least not to the extent that is needed in the enterprise data warehouse. When a large amount of historical data starts to stack up in the ERP environment, the ERP environment is usually purged, or the data is archived to a remote storage facility. For example, suppose an enterprise data warehouse needs to be loaded with five years of historical data while the ERP holds at the most, six months worth of detail data. As long as the corporation is satisfied with collecting a historical set of data as time passes, then there is no problem with ERP as a source for data warehouse data. But when the enterprise data warehouse needs to go back in time and bring in historical data that has not been previously collected and saved by the ERP, then using the ERP environment as a primary source for the data warehouse is not a viable option.

Metadata in the ERP is another consideration when building a data warehouse is in the ERP environment. As the metadata passes from the ERP to the data warehouse environment, the metadata must be moved and transformed into the format and structure required by the data warehouse infrastructure. There is a significant difference between operational metadata and DSS/DW metadata. Operational metadata is primarily for the developer and programmer. DSS metadata is primarily for the end user. The metadata that exists in the ERP application's database must be converted, and such a conversion is not always easy or uncomplicated, and requires experienced data administrators and users to collaborate in the effort.

Mr. Inmon suggests some guidelines for using the ERP database as a source for a data warehouse. They would include the existence of a solid interface that pulls data from the ERP environment to the data warehouse environment. The ERP to enterprise data warehouse interface needs to:

- be easy to use
- enable the access of ERP data
- capture the meaning of the data that is being transported into the data warehouse
- be aware of restrictions within the ERP that might exist when it comes to the accessing of ERP data
- be aware of referential integrity
- be aware of hierarchical relationship
- be aware of logically defined - implicit - relationships
- be aware of application conventions
- be aware of any structures of data supported by the ERP
- be efficient in accessing ERP data, supporting -
    - direct data movement
    - change data capture
- be supportive of timely access of ERP data
- understand the format of data

(Taken from "Data Warehousing and ERP", a white paper by Wm. H. Inmon, Kiva Productions, LLC, 1999)

In summary, the development of data warehouses and the emergence of ERP as factors in the information systems explosion must be addressed and resolved by experienced information systems professionals with a clear understanding of the challenges each environment poses. Integrating ERP data into a data warehouse can lead to a superior source of data for analysis and decision-making if the data is formatted for query and reporting, and if the ERP environment is coordinated with the decision support needs of the organization. To ignore the wealth of data and information that is available from an ERP is to ignore a valuable corporate resource, one that can serve as a foundation for a superior data warehouse.

## 3. THE RELATION BETWEEN DATA WAREHOUSE AND BUSINESS INTELLIGENCE

The number of business intelligence (BI) solutions appearing in the marketplace is steadily increasing. Most of these solutions still employ the services of a traditional enterprise data warehouse, but an increasing number do not. In some operational BI applications, for example, event data volumes and/or the need for fast action times may prevent the data from being persisted in a data warehouse before it is analyzed. These latter applications do not replace the enterprise data warehouse, but often work in conjunction with it – the results from the analytical processing may be stored in the enterprise data warehouse, for example.

As the BI industry evolves, so too does the role of the traditional enterprise data warehouse. The title of this section is deliberately provocative. The objective is to encourage BI professionals to consider the role of the data warehouse in new BI projects. In the past, the enterprise data warehouse has been the cornerstone for such projects, but WE believe in many situations this is no longer true.

That same implementation, now a decade and a half removed, even then challenged the concept of either historical or strategic — and it was telecom. The biggest distinction was that it was informational vs. operational — but the implementation was clearly tactical vs. strategic and needed to be as near real-time as could push it — they used it to evaluate the data (based on algorithms honed from data mining) and more appropriately segment call center leads for highest-value use — given that regulation prevented calling a household more than once in a three month period, the goal was to call them for the 'right' (highest value) campaign.

What they call it all is less important than what the terms help their accomplish in reaching a common understanding. Even as they were in the throes of 'making up' what data warehousing was, they were challenging its precepts

## 4. DATA WAREHOUSE DESIGN STRATEGIES

To build an effective data warehouse, it is important to understand data warehouse design principles. If data warehouse is not built correctly, it can run into a number of different problems. Once designed data warehouse that is user friendly,it will next want to look at operational efficiency. Once the data warehouse has been created, it should be able to carry out operations quickly. In addition to this, it should not have errors or other technical problems. When errors or technical problems do occur, they should be simple to fix. Another thing that will want to look at is the cost involved with supporting the system and will want to keep these costs low as much as possible.

The design principles that have been discussed in this paper so far are more related to business than information technology. However, there are a number of IT design principles that will want to follow. One of these is scalability. This is a problem that many data warehouse designers run into. The best way to deal with this issue is to create a data warehouse that is scalable from the beginning. Design it in a way which will allow it to support expansions or upgrades. It should be able to adapt it to a number of different business situations. The best data warehouses are those which are scalable.

The data warehouse that designed should fall under the guidelines of information technology standards. Every tool that use to build the data warehouse should work well with IT standards. You will want to make sure it is designed in a way that makes it easier for workers to use. While following the guidelines in this paper won't allow to always be successful, it will greatly tip the odds in the favor. No matter how well designed data warehouse is, it will always run into problems. However, following the right principles will make the problems easier to recognize and solve.